\documentclass[sn-standardnature]{sn-jnl}

\jyear{2024}%

\theoremstyle{thmstyleone}%
\theoremstyle{thmstyletwo}%

\theoremstyle{thmstylethree}%

\begin{document}

\title[How to write competitive proposals and job applications]{How to write competitive proposals and job applications}


\author*[1,2]{\fnm{Johan H.} \sur{Knapen}}\email{johan.knapen@iac.es}

\author[3]{\fnm{Henri M.J.} \sur{Boffin}}\email{hboffin@eso.org}

\author[4]{\fnm{Nushkia} \sur{Chamba}}\email{nushkia.chamba@nasa.gov}

\author[5]{\fnm{Natashya} \sur{Chamba}}\email{natashya.c@nsbm.ac.lk}

\affil*[1]{\orgname{Instituto de Astrof\'\i sica de Canarias}, \orgaddress{\street{Calle V\'\i a L\'actea S/N}, \city{La Laguna}, \postcode{E-38205}, \state{}\country{Spain}}}

\affil[2]{\orgdiv{Departamento de Astrof\'\i sica}, \orgname{Universidad de La Laguna}, \orgaddress{\city{La Laguna}, \postcode{E-38200}, \state{}\country{Spain}}}

\affil[3]{\orgname{European Southern Observatory}, \orgaddress{\street{Karl-Schwarzschild-str. 2}, \city{Garching}, \postcode{DE-85748}, \state{}\country{Germany}}}

\affil[4]{\orgname{NASA Ames Research Center}, \orgdiv{Space Science and Astrobiology Division}, \orgaddress{\street{M.S. 245-6}, \city{Moffett Field}, \postcode{CA 94035}, \state{}\country{USA}}}

\affil[5]{\orgname{NSBM Green University}, \orgaddress{\street{Pitipana}, \city{Homagama}, \postcode{10206}, \state{}\country{Sri Lanka}}}


\abstract{Writing proposals and job applications is arguably one of the most important tasks in the career of a scientist. The proposed ideas must be scientifically compelling, but how a proposal is planned, written, and presented can make an enormous difference. This Perspective is the third in a series aimed at training the writing skills of professional astronomers. In the first two papers we concentrated on the writing of papers, here we concentrate on how proposals and job applications can be optimally written and presented. We discuss how to select where to propose or apply, how to optimise your writing, and add notes on the potential use of artificial intelligence tools. This guide is aimed primarily at more junior researchers, but we hope that our observations and suggestions may also be helpful for more experienced applicants, as well as for reviewers and funding agencies.

This is a preliminary version of a paper published in Nature Astronomy, 9, 951 (2025). Please read the paper there, or download the published version from \href{https://rdcu.be/evDCt}{https://rdcu.be/evDCt} . }




\maketitle

\section{Introduction}\label{sec1}

Writing proposals forms a fundamental part of modern science. Whether you request observing or supercomputing time, ask for a grant to fund students, post-docs, or equipment, or if you apply for a position, you will have to write a competitive proposal, which is typically peer-reviewed. Proposals are often solicited, submitted after careful writing and discussion within a team, then evaluated by a panel of fellow scientists. Proposals are thus among the most important documents that you will have to write.
Indeed, without successful proposals and job applications, it is hard to conceive that you will have a career as an independent scientist. 

This Perspective is the third in a series on scientific writing for professional astronomers. Many aspects discussed in Papers I \citep{Chamba22} and II \citep{Knapen22} on how to plan, write and organise your paper in astronomy are also fully applicable to proposal writing and shall be referenced where appropriate. Additional background material can be found \href{https://writebetterproposals.org}{here}, \href{https://www.ogrants.org}{here}, \href{https://www.nhfp-equity.org/resources-for-applicants}{here}, \href{https://www.nature.com/articles/d41586-022-02958-4}{here}, \href{https://www.nature.com/articles/d41586-023-01881-6}{here}, \href{https://www.nature.com/collections/iihihcfghe}{here}, and \href{https://www.nature.com/collections/bihhaafahc}{here}.

In this Paper~III, we address points which are specific to writing proposals, including job applications. We will share our experience in writing, handling and evaluating proposals in astronomy, and add thoughts about modern developments, including the possible use of artificial intelligence (AI) large language models (LLMs) like ChatGPT. Just like Papers I and II, this paper is aimed at beginning writers in professional astronomy, but should also be useful to more senior astronomers and scientists in adjacent fields.

\section{Proposal basics}

Allocation of resources or jobs is competitive. At the European Southern Observatory (ESO), for example, some 900 proposals are received every six months to request observing time on the La Silla Paranal Observatory's telescopes. Oversubscription at these facilities, as on telescopes such as ALMA, the HST or the JWST, varies from factors of 5 to over 15. The numbers are as bad for many grants and jobs---the success rate for European Union ERC or MSCA calls is often below 10\%, and fellowships and permanent positions may attract dozens of candidates. 

It is essential to carefully study the call for proposals or advert, and the criteria for evaluation that should be (but often are not) published along with it. Is the call really for you? Do you fulfil all the criteria? Is it worth investing time in writing something of quality, given the expected success rate? The answer to these questions will depend on the kind of proposal that is required (is it 2- or 30-page long? does it need extensive preparation or can you recycle material? can you provide all requested aspects, like a Gantt Chart or evidence of previous experience?), the potential gain (some additional telescope time or a career-making grant or permanent position?), and whether you are preparing the proposal alone or within a team. Usually it is the latter, in which case team members should be invited as early as feasible so they have time to comment and you have time to fully use their expertise.

Using an indiscriminate approach to respond to many calls or advertisements is usually not a good idea. Low success rates mean that only the most competitive or relevant proposals are selected, not necessarily the lucky ones. As the effort needed to produce a competitive funding proposal can be as large as that involved in writing a research paper, choosing your target call is important, unless material can easily be recycled (e.g., in job applications) or proposals are quick to produce (e.g., some telescope time or travel grant applications). In our experience, ambitious proposals can take 3-6 months to write, of which 1-2 months essentially full-time, and may only get approved after one or more resubmissions.

Specific sections which you are not familiar with may require extra effort. Examples are `impact' in many EU proposals, or custom statements on teaching, diversity, or open access publishing. What can also take time is explanatory work in terms of pilot studies or developing links to new partners in industry or elsewhere. Reviewers are more likely to be convinced by the technical details and methodology of your proposal if you also demonstrate their feasibility. For first time writers, this step may well be time consuming as it will often require using dedicated tools as well as crafting clear figures and graphics. Find a balance which will convince the expert reviewer but not put off her non-expert colleague by a flood of numbers and details.

Once you decide to respond to a certain call, define the scientific idea that must support your proposal. Try to summarise it in one sentence, then in a paragraph. Use that to derive the specific goals or objectives. Then construct your proposal from there. Identify your approach by studying the call for proposals and plan how and when to write each (sub-)section. Consider who can help you, from trusted scientific colleagues to grants office staff or external consultants, and enlist their help as early on as feasible. If the proposal language (normally English) is not your own, or if sections (e.g., abstract) are needed in a language that you do not dominate, plan for adequate help and review. Paper II has useful hints in this regard, in particular on writing mechanics. Study the length of sections and aim for a balance that is right in response to the call, rather than being the easiest to write for you. Plan when to have drafts ready for review. Proposals normally have deadlines but these are random dates---consider setting your own deadline well ahead of the official one to create time for review by yourself and others. 

\section{What makes a proposal successful?}

A proposal can be seen as a {\href{https://www.inc.com/geoffrey-james/how-to-write-a-winning-proposal.html}{sales}} tool, rather than an information package. Like in business, your proposal should convince the panel members (clients) that they should approve your proposal (buy your product) and not those of others.

A proposal requires you to be convincing, but also to the point. Given the workload of your reviewers, who will do their evaluating tasks on top of their usual duties, your proposal must be captivating, allowing the reviewer to understand immediately what you want to do, and enthuse them. Your proposal also needs to be concise. Most proposals have very strict constraints: maximum number of pages, specific sections, fonts and font sizes, margins, etc. Ensure that your proposal is compliant.

The committee in charge of evaluating the proposal is composed of peers. For large observational facilities, such as at ESO, ALMA, HST, or JWST, there will be sub-panels that consist of experts in a broad area, but in most national time allocation committees or in hiring committees, there is only one group. Even if funding proposals are seen by fellow astronomers (they may not---find out in advance who will judge your proposal!), it is unlikely that more than one, if any, will actually work in your specific area. Your proposal must therefore be understandable for a non-expert. It must be explicit, and you should never assume that the panel will work out what you meant. Explain your points in general terms, but avoid being seen as condescending to that one panel member who actually {\it is} that world expert on your exact science (see Sect. 2.5 in Paper I, on the basics of writing style).

A successful proposal will likely combine several aspects. First, you need to convince reviewers, even if they are not within your field of expertise, that your proposal is scientifically relevant and exciting. Second, the outcome will appear fundamental rather than incremental, and lead to substantial progress. Third, the author is the key point in making this possible---thanks to some unique skills or a unique idea; an important point to convince a committee that they should select this proposal and not another one. Fourth, the methodology should be sound and the timeline convincing. 
Fifth, and most important, a compelling proposal is presented as a story that is nice to read and easy to follow, and that will make an evaluator enthusiastic (or even `feel warm inside'). The final goal of the proposal is that the reviewer wants to know the end to your story and will therefore recommend approving your application rather than those by others.

The Spitzer Space Telescope Science Center provides {\href{https://slideplayer.com/slide/8168642/}{guidance}} to potential applicants, including: 

\begin{quote}
``Good proposals include some background on the subject you are studying, in particular why anyone not in your specific field should care. Then you can explain what exactly you want to do, and why it will solve every problem left in astronomy and find a cure for the common cold. Adding good figures and tables almost always makes a proposal stronger and easier to understand for the reviewers."
\end{quote}

Although an exaggeration, it illustrates the point of a successful proposal: it needs to be ambitious though not bragging, while providing all relevant information (see Sect. 2.1 in Paper I which discuses how to link the scientific background with the gap in the literature that you aim to address). All this has a clear implication: writing a good proposal is not an easy task, and it takes time. 

\section{Writing a competitive proposal}

The same rules of good writing for scientific articles also apply to proposals \citep{Chamba22,Knapen22}. The structure of the hourglass \citep[see, e.g.,][]{Ivanochko21} is important: because the reviewer is not necessarily an expert in your field, you start with the general picture as to why your research area is of high interest and what exactly is the problem, with an emphasis on explaining why it needs attention now. Then you funnel to the middle part of the hourglass, describing the details of what you will study, and how you will solve the stated problem. Towards the end of the proposal, you open up again and state how the field will have progressed once you have done the work you propose. It is essential that you provide an up-to-date, comprehensive bibliography in your proposal, with both historical and the latest references. Making these references into clickable links makes it easier for reviewers who read your proposal online.

Similarly to how you draft your papers, when writing your proposal, we recommend that you start by creating an outline of what you will discuss, including sections, sub-sections, paragraphs, etc. As in papers, make the most of your section titles by avoiding generic ones like `Methods'.Then write a quick first but complete draft. Spend the remaining time on revising it. It is often useful to play the role of `critical reviewer' to your own proposals as you revise it. If you are afraid to delete a piece of well-written text, copy and move it elsewhere. N(u)C often has a `JUNK.txt' file for various pieces of text and well crafted sentences she did not want to completely get rid of and may find useful in the future. 

Remove everything that is not relevant or needed, from single words to sections. Add missing information. Avoid rambling, in particular in sections you are not expert on (for those, ask for advice, or be honest). Make your sentences as clear as possible. Short sentences may be clearer, but your own cultural background may push you towards longer ones (e.g., German or Spanish tend to require more elaborate sentences). Iterate several times! Once you have something that seems complete, show it to colleagues, including those who do not work in your field, and ask for their honest feedback. Then, revise again. In all this, make sure the deadline is far enough in the future that there is ample time for each revision step.

Four parts of your proposal are particularly important: the title, the abstract, the opening paragraph, and the conclusion. For instance, proposals may be allocated to reviewers by a panel chair who will not read your entire proposal but will base their choice on these items. Also, a proposal may well be read in detail by two referees or rapporteurs, who will report on it to the wider panel of people who may not read your entire proposal in detail, but will base their own opinion on the four items mentioned above, plus additional ones like figures.

The current writers do not agree on whether it is best to draft your title and abstract before anything else (which capture the main message, allowing you to construct your proposal around the main points) or at the very end (when you know what the proposal says). As with papers, there are few hard rules, so choose what works for you. But ensure that your title is catchy, relevant, short, and with as many keywords as feasible. For instance, in a JWST proposal there is no need to include `JWST' in the title. `Unveiling' is too vague to be useful, choose a better word. If your object of interest is, say, young, a title often does not need to detail exactly how young. Acronyms, jargon and object names are usually best avoided in titles.

Similarly, the abstract should summarise the excitement in one paragraph and rather not contain jargon or assume specialist knowledge. Here also, try to follow the hourglass structure, from the big picture, via the details of the problem the proposal will address and your solution, to the expected goal and outcome and how this will change the field. Managing to do this in the space constraints is not simple, and you may see yourself writing it over and over again. 

A typical mistake of many proposers is to try to pack as much text as possible into the available space. This will make the reading very hard, frightening any reviewer. Try to liven up large blocks of text, in particular in long proposals, by carefully including and distributing white space, nice figures and tables, text boxes, highlight colours, etc.---though without making your proposal look like the graphical equivalent of a Christmas tree. Another tip is to concentrate on one or two scientific questions rather than many related issues.

Kuchner \cite{Kuchner_Marketing} suggests that a perfect proposal includes three different kinds of figures: a beautiful butterfly figure, a family portrait, or a before/after figure. The beautiful butterfly figure is a figure whose purpose is to grab the reader's attention and set the stage. It could be a nice image of a galaxy or a nebula, or an image of a simulation. Try to personalise this figure, and avoid using that picture from an Annual Reviews paper which is so great that everyone in your field has been showing it for a decade. 

The family portrait is a figure where you present the state of the art of the field. It is your best chance to show that you are familiar with the current state of your research area and to acknowledge the work done by all your peers. It could for example show a mass versus orbital period diagram for all currently characterised exoplanets, highlighting some of the most relevant examples and indicating where your proposal will solve an outstanding issue. 

Finally, the before/after figure is what we know from advertisements. Show the current state of research and next to it, how it would improve should the proposal be approved. The more striking the difference, the more chance you have for the proposal to be accepted. This could for example illustrate a 1-sigma contour of a variable in a given plane and how this would dramatically shrink should more observations be available. 

Only in the largest proposals is there generally space to include all of these. In smaller proposals, a before/after figure may be your best bet as this is what will convince the reviewer. Also consider your panel---if your proposal is judged by experts they do not need a picture of a galaxy, whereas that same picture may strike a chord with reviewers from, e.g., theoretical physics or social or life sciences. Whether your panel contains subject specialists or not, ensure that your figures are clear, legible, and that all aspects are explained. Often a new version of a published figure (quoted as `based on Author et al.') is preferable to a relevant but cluttered one.

\subsection{Specific writing tips}

{\bf Avoid generic statements.} A sentence like: {\it `The study of galaxy evolution/ GRBs/ the Moon is one of the key fields of modern astrophysics.'} will not tell the reviewer anything new. Instead, start with a sentence that will trigger their interest and grab their attention \citep{SSRC95}. This is what Hollywood does in the movies: {\it `a  great film hooks viewers through an early scene that raises an exciting question'} \citep{Hartmann23}. Effective communication is about knowing your audience; likewise, you can write a better proposal if you understand what the reviewer knows about your problem and what you need to tell them. Your opening paragraph is the key to explain the problem at hand. 

{\bf Keep it simple.} Write concise and clear sentences. Carefully review whether any long sentences are really needed, and really work. Remember that your native language may push you towards longer sentences. Reviewers are busy and will appreciate that your text is agreeable to read. Limit jargon and abbreviations. Do not use contractions or exclamation points. Avoid typos. Use elements like text boxes or figures to break large blocks of text. Short sentences can be effective.

{\bf Make every sentence count.} In proposals, it is even more important than in most papers to remove dead wood and filler words. You can for example replace  `the majority of' by `most'. Steer clear of words such as `unique' (can you really prove it?), `maximise' (be quantitative instead), or `I believe' or `I think' (believing is a matter for the faithful---we want facts). More tips are in \cite{Knapen22}. 

{\bf Project enthusiasm.} You can use words such as `stunning', `remarkable' or `awesome' but there is a fine line between enthusiasm and boasting. Feedback from your peers will help you avoid overdoing it, in particular if your language is not that of the reviewers. 

{\bf Beware of biases.} Alarmingly, gender and culture often influence how proposals are written, and how successful they are. For instance, \cite{Lerchen18} report that male first authors are more likely to present research findings positively than female first authors. To complicate matters more, however, \cite{Valian23} stated that {\it `in anonymous proposals, women should not refrain from using positive words to emphasise their work, but,'} [because of biases], {\it `in non-anonymous proposals, they need to refrain to do so as people don't like women who brag'}. Another strong reason, in our opinion, to use anonymous peer review, train all reviewers, and make panels use well-defined criteria rather than value judgements like `high-quality science', `importance for the wider field', or `excellence'. We point out that biases tend to be persistent, for instance, \cite{Primas24} conclude from a detailed analysis of the impact of dual anonymous peer review on ESO telescope time allocations that this has not cancelled the gender gap in the success rate of proposals led by male or female researchers.

Cultural and regional biases also exist. For example, \cite{ALMA23} analyse the ranks of ALMA proposals as a function of the region of the principal investigator (PI). They found that proposals from Chile, East Asia, and other regions continue to have poorer proposal ranks relative to PIs from Europe and North America even after randomising the investigator list (as was done in ALMA Cycle 7) and the introduction of dual-anonymous review (in ALMA Cycle 8), although the ranks of Chilean proposals appear to have improved slightly. The cause of this discrepancy is not clear, but it may be due to a different use of language, with North American and European PIs using more enthusiastic language. 

{\bf Avoid using negative expressions.} These leave a more lasting impression than positive ones---something called negativity bias \citep{Peeters90}. Moreover, negative sentences are also often more difficult to understand. If you do need to share bad news or a caveat, try not to do it in a first or final sentence of a paragraph.

\section{Job applications}

An important kind of proposals are job applications. Get spelling and grammar right throughout (one of us remembers once seeing an application with the applicant's name misspelled---it did not succeed). Read the job advertisement carefully to know if you are asked to submit a `research statement' (mostly a summary of past research) or a `research plan' (a proposal for research projects to be undertaken during the job if hired). In particular the latter follows the principles of a proposal as indicated above---within the requested length and format, and ideally including the three kinds of figures. 

In addition, you are usually asked to add a motivation (or cover) letter and a curriculum vitae (CV). Check out examples of CVs on the web, and ask peers for feedback on your chosen format. The motivation letter is important, and should be tailored to the position you are applying for. As you recycle your application and letter, ensure to refer to the right position, institution and people. 

Scientific independence is sometimes a criterion, often assessed on the basis of papers of the applicant produced without their supervisors. Evidence of soft skills or management or outreach activity can also be a selection criterion, so keep this in mind when chosing your activities. Help in conference or seminar organization, for instance, is a useful addition to your CV.

Always try to address the cover letter to a person, find out who chairs the panel, for instance. Do not assume they are a man, or a married woman (Mrs.); referring to someone as Professor or Dr is always safe, even if they turn out to be an administrative assistant, while the opposite is usually not seen in such a positive light. Add names of people who you would be interested in working with, especially if you know they will be involved in the selection. Panel members are people, and we all like to see our work and interests respected and referred to. In general, the recruiters need to see that you made an effort to learn about their group or institute, and how you will fit in. 

For job applications you generally need to provide the names of a number of referees. Aim to select people that know you well and who will take the time to write detailed and personalised recommendation letters---if you are aiming for a PhD position, ask your MSc advisor, for a postdoc position, your PhD advisor. In particular for (independent) fellowships, that a referee knows how you work and can vouch that you can deliver certain goals may be more important than a `big name' writing a reference. Provide your referees, well ahead in time, with all the necessary information: what the job is about; who to send the letter to; what is your research plan; etc. They are busy people and writing a reference letter is a serious and time-consuming job, so your referees will be happy if you draft paragraphs or sentences for them, e.g., summarising your main papers and why they are so important. Keep your referees informed, including of the outcome of the process.

A profoundly unpleasant feature of the widespread use of references in our field is that a vital part of a job application is not under the applicant's control, and usually not even seen by them. It is a fact that some people write negative references. Many more think they are positive but their letters are not perceived as such due to how they write them. To spread your risk, we recommend not to use the same reviewers for all your applications---ask four referees for instance and then select changing sets of two or three. 

If for any reason you cannot or would rather not ask your direct supervisor or employer to write a reference, try to ask one of your other referees to refer to this in terms which are positive for the applicant. As this is not always feasible, it is vital that panel members always assume that a missing key reference is due to specific circumstances (which can, unfortunately, include unpleasant ones like harassment or abuse), rather than a weakness of the applicant.

\section{Intelligent use of artificial intelligence}

For most of us, including all current authors, English is not our mother tongue and writing a competitive proposal in English is thus a double challenge. One way to cope with this is to use AI tools. For example, the {\tt Deepl} translation software can help if you find it easier to draft sentences or sections in your mother tongue. Chatbots based on large language models (LLMs), such as {\tt chatGPT}, {\tt Claude}, or {\tt Gemini}, are remarkably powerful in producing text which is well constructed in terms of spelling and grammar.

Using LLMs as a tool to enhance your language skills and perfect your writing style is very interesting, as is asking them for inspiration if you are stuck with, e.g., a title, a specific sentence, or even a paragraph. You can ask LLMs to correct and proofread your own English, to find errors and inconsistencies in spelling, grammar, or even logic. In this way, you can use them to improve your writing whilst maintaining your style and voice, as opposed to using LLMs to write from scratch. You can also use it as an English training tool by asking it to analyse your writing.

N(a)C frequently incorporates AI into her lectures. She teaches students how to frame a prompt into an LLM that yields the best results or how to use its audio mode as a conversational partner \citep{kruse23}. To summarise complex information, her students need to state the topic, then explicitly ask the LLM to break down the topic into smaller, easier-to-understand parts, with real-life examples and analogies for better comprehension. The more descriptive and detailed your prompt, the better the output from the LLM. In fact, writing prompts has garnered so much attention that it is now officially a job role in many tech firms. If you want a LLM to analyse part of your proposal and give you effective feedback, clearly define the parameters and the guidelines of the proposal so that the LLM can align to this. When crafting paragraphs for your proposal, you can provide the LLM with samples of your own writing prior to the prompt, to guide the LLM to generate text similar to your style of writing.

The use of LLMs can, however, limit your learning, and can lead to ethical problems including plagiarism. Sharing proposals with these tools, especially when reviewing proposals, is often against the rules of funding organisations, and may mean that they will become public---check in detail before using any AI tools. Using LLMs for certain sections or paragraphs can lead to different writing styles across your proposal, which reviewers will not appreciate. Another key point is that LLMs, by definition, predict from historical use of language, where a competitive proposal should present and explore new ideas. 

When asking LLMs to write as well as evaluate telescope time proposals, \cite{Jerabkova24} found that chatGPT-adjusted proposals receive lower grades than the original proposals. When evaluating, chatGPT is very good at summarising but less good at identifying weaknesses, and tends to prefer proposals written by itself. Jerabkova et al. \cite{Jerabkova24} confirm that LLMs tend to be verbose, without flavour, may invent facts and references (see also \citep{Kabir23}), and in any case need very specific instructions to generate useful output. Fouesnau et al. \citep{Fouesneau24} provide a detailed look at the possible role of LLMs in astronomy, while much information is also available in the series of {\href{https://youtube.com/playlist?list=PLLLg88mtP9s7olsC5-JEAnGKvKoKYBVMv}{ESOGPT24}} talks. 

Always treat the outcome of LLMs as you would any draft, and edit, correct, proof-read, and check the results, and ask others to read your corrected version. Look out for words which you would not normally use or which are not common in our field. Our recommendation is to embrace these new AI tools to improve your writing or your efficiency, but to remain very critical of their output, in particular for such important and unique documents as proposals.

\section{Summary remarks}

By preparing well and training yourself in writing proposals, you can be successful even though most proposals and applications fail. How to achieve that? First of all, develop good scientific ideas (which we cannot help you with). Then, study how to write, using, e.g., Papers 1 and 2 in our series, follow those guidelines which you think will work for you, choose critically and realistically which calls or advertisements to target, discuss your proposals with others (peers, supervisors, collaborators, successful past applicants, commercial consultants if you can find funding), and analyse feedback on earlier submissions.

Success is almost always the result of hard work and dogged determination (for instance, JHK's currently funded EU Doctoral Network project\footnote{\url{https://research.iac.es/proyecto/educado/}} was the result of the proposal failing twice before, and only scraping through the selection in the third round). Of course the news is dominated by success stories of major grants or awards, but even the most successful among us must deal with more disappointments than successes. Without exception, every single scientist could publish a `CV of failures' \cite{Stefan10}, and it would be good if more admitted to this.

We finish with some final recommendations for re-submissions---since most proposals and job applications, unlike papers, can be re-used and improved with time. 
\begin{itemize}
\item Never take a rejection personally: it is of the actual proposal submitted at a given time, not of your person. The same proposal that was brutally rejected one day may well be approved with very positive comments when re-submitted. 
\item Critically read the feedback from the reviewers and use it to improve your application, preferentially after discussing it with colleagues. Ask for feedback if it is not provided by default.
\item Go through the proposal and wherever possible make it clearer, shorter, more compelling. 
\item Ensure that there are no errors, and remove any vague or ambiguous statements. 
\item Strengthen your science case, e.g., by using archive data or simulations to show that the proposed strategy works. 
\item Publish on-going projects, so that the committee will not doubt your productivity. 
\item Increase your visibility by giving talks and by networking during conferences. Your future referee may be one of the attendants or the person you just talked to.
\item Finally, analyse why a certain proposal or application, however modest, is successful, and celebrate those successes. Good luck!
\end{itemize}
\backmatter

\bmhead{Acknowledgments}

We thank our colleagues for providing input and suggestions at various stages of the preparation of this manuscript. In particular, we thank Drs S\'ebastien Comer\'on and Junais for providing detailed comments.

\section*{Declarations}

{\bf Funding} Co-funded by the European Union (MSCA EDUCADO, GA 101119830, and Widening Participation, ExGal-Twin, GA 101158446). Views and opinions expressed are however those of the author(s) only and do not necessarily reflect those of their employers, nor of the European Union. Neither the European Union nor the granting authority can be held responsible for them. JHK acknowledges grant PID2022-136505NB-I00 funded by MCIN/AEI/10.13039/501100011033 and EU, ERDF.

\bmhead{Authors' contributions} 

JHK and HB drafted the manuscript combining their years of experience as professors, lecturers, reviewers, and application process managers. N(u)C and N(a)C helped develop the final text from the perspective of a postdoc in astronomy and an English language lecturer using AI in a developing country. 

\bmhead{Competing Interests} 

The authors declare no competing interests.



\end{document}